\documentclass[letter]{aa} 

\usepackage[varg]{txfonts}
\usepackage{siunitx}
\usepackage[version=4]{mhchem}
\usepackage{multirow}
\usepackage{graphicx}
\usepackage[caption=false]{subfig}
\usepackage[strict]{changepage}
\usepackage{xcolor}
\usepackage{ulem}
\usepackage[colorlinks=true,citecolor=blue]{hyperref}
\usepackage{float}
\usepackage{amssymb}
\usepackage{amsmath}
\usepackage{lscape}
\usepackage{mathtools}
\usepackage{natbib}
\usepackage{pdflscape}
\usepackage{gensymb}

\defcitealias{zhou2023}{Paper~I}   

\DeclareSIUnit\jansky{Jy}

\newcolumntype{L}[1]{>{\raggedright\let\newline\\\arraybackslash\hspace{0pt}}m{#1}}
\newcolumntype{C}[1]{>{\centering\let\newline\\\arraybackslash\hspace{0pt}}m{#1}}
\newcolumntype{R}[1]{>{\raggedleft\let\newline\\\arraybackslash\hspace{0pt}}m{#1}}

\begin{document}

\title{Inversely synthesizing the core mass function of high-mass star-forming regions from the canonical initial mass function}
\author{J. W. Zhou\inst{\ref{inst1}} 
\and Pavel Kroupa\inst{\ref{inst2}} \fnmsep \inst{\ref{inst3}}
\and Sami Dib \inst{\ref{inst4}}
}
\institute{
Max-Planck-Institut f\"{u}r Radioastronomie, Auf dem H\"{u}gel 69, 53121 Bonn, Germany \label{inst1} \\
\email{jwzhou@mpifr-bonn.mpg.de}
\and
Helmholtz-Institut f{\"u}r Strahlen- und Kernphysik (HISKP), Universität Bonn, Nussallee 14–16, 53115 Bonn, Germany \label{inst2}\\
\email{pkroupa@uni-bonn.de}
\and
Charles University in Prague, Faculty of Mathematics and Physics, Astronomical Institute, V Hole{\v s}ovi{\v c}k{\'a}ch 2, CZ-180 00 Praha 8, Czech Republic \label{inst3}
\and
Max-Planck-Institut f\"{u}r Astronomie, K\"{o}nigstuhl 17, 69117, Heidelberg, Germany
\label{inst4}\\
\email{dib@mpia.de}
}

\date{Accepted XXX. Received YYY; in original form ZZZ}

\abstract
{Many studies have revealed that the core mass function (CMF) in high-mass star-forming regions is top-heavy. In this work, we start from the canonical initial mass function (IMF) to inversely synthesize the observed CMFs of high-mass star formation regions, taking into account variations in multiplicity and mass conversion efficiency from core to star ($\epsilon_{\rm core}$). To match the observed CMFs, cores of different masses should have varying $\epsilon_{\rm core}$, with $\epsilon_{\rm core}$ increasing as the core mass decreases. However, the multiplicity fraction does not affect the synthesized CMFs. To accurately fit the high-mass end of the CMF, it is essential to determine whether the CMF shows a slope transition from the low-mass end to the high-mass one. If the CMF truly undergoes a slope transition but observational biases obscure it, leading to a combined fit with a shallower slope, this could artificially create a top-heavy CMF.}


\titlerunning{from IMF to CMF}
\authorrunning{J. W. Zhou, Pavel Kroupa}

\maketitle

\section{Introduction}
Giant molecular clouds (GMCs) are widely recognized as the primary gas reservoirs that fuel star formation and serve as the birthplaces of nearly all stars. A dense core is a localized fragment or overdensity within a molecular cloud, representing a local minimum in gravitational potential and serving as the precursor to an individual star or stellar system \citep{Andre2014}. A prestellar core is a dense core that remains starless while being gravitationally bound.
In star-forming regions within the solar neighborhood that produce solar-type stars, the prestellar core mass function (CMF) closely resembles the initial mass function (IMF) \citep{Alves2007-462, Enoch2008-684, Andre2014, Konyves2015-584, DiFrancesco2020-904}. This similarity has led to the hypothesis that the IMF's shape may be directly inherited from the CMF. However, the Milky Way includes both low-mass and high-mass star-forming regions. A thorough analysis of large statistical samples of massive protoclusters is essential to observationally determine whether the origin of the IMF is truly independent of cloud properties.
Thanks to observations with the Atacama Large Millimeter/submillimeter Array (ALMA), numerous studies have now examined CMFs in high-mass star-forming regions \citep{Motte2018-2,Liu2018-862,Cheng2018-853,Sanhueza2019-886,Lu2020-894,Sadaghiani2020-635,O'Neill2021-916,Pouteau2022-664,Nony2023-674,Louvet2024-690}, revealing that the CMFs in these regions are flatter than the ones in nearby, low-mass star-forming regions. The excess of high-mass cores in these regions indicates a top-heavy CMF, which may challenge the universality of the IMF.
In the literature, the CMF fitting for high-mass star-forming regions typically does not differentiate between core types (such as prestellar or protostellar). However, the differentiation was done in \citet{Sanhueza2019-886,Nony2023-674}. \citet{Sanhueza2019-886} derived a power law index of -2.17$\pm$0.10 for the prestellar CMF, which is slightly shallower than the Salpeter IMF. In \citet{Nony2023-674}, on the other hand, the prestellar CMF has a power law index of -2.46$\pm$0.15. The protostellar CMF in \citet{Nony2023-674} is significantly shallower than the Salpeter value, with an index of -1.96$\pm$0.09. As is shown in Sec.\ref{o-cmf}, the observed CMF of high-mass star-forming regions may not be robust.

It is now widely accepted that the majority of stars both form and exist on the main sequence with at least one stellar companion; “single stars are the exception, not the rule” (see \citet{Offner2023-534} and references therein). \citet{Goodwin2005-439} argued that the short ($<10^5\,$yr) dynamical decay times of nonhierarchical multiple systems imply that most stars must form in binary systems or dynamically stable hierarchical higher-order systems to explain the observed very high fraction of binaries in about $1\,$Myr old star-forming regions. 
Multiple star systems are formed during the earliest stages of star formation. Observations have offered evidence that the multiplicity fraction of protostars is higher than that of pre-main-sequence and main-sequence stars \citep{Looney2000-529, Chen2013-768}, indicating that most systems initially form as multiples. Observations of multiplicity in the youngest stellar systems are key to understanding their origins \citep{Izquierdo2018-478,Olguin2021-909,Olguin2022-929,Tobin2022-925,Li2024-8}. As is shown in Fig.1 of \citet{Offner2023-534}, the likelihood of massive protostars forming as multiple systems may approach nearly 100\%. 

The multiplicity of protostars is tightly correlated with the origin of the IMF or the transformation from the  CMF to the IMF. Despite the high multiplicity fraction among protostars, it is often assumed that a single core will produce only one star. 
A core at low resolution may actually consist of multiple smaller-scale cores, which should form a multiple system rather than a single star. Assuming a one-to-one mapping between cores and stars artificially shifts the cores to the higher-mass range, potentially making the CMF appear top-heavy.
Apart from the multiplicity, the mass conversion efficiency from core to star ($\epsilon_{\rm core}$) will also have a significant impact on the transformation between the CMF and the IMF. For the cores in star-forming regions within the solar neighborhood, the values of $\epsilon_{\rm core}$ are $\approx$ 20$-$40\% \citep{Alves2007-462,Konyves2010-518,Konyves2015-584,Megeath2016-151}. 
In this work, we start from the canonical IMF to inversely synthesize the observed CMF of high-mass star formation regions, taking into account variations in multiplicity and $\epsilon_{\rm core}$.

\section{Sample}

For three high-mass star-forming regions, W43-MM1\&MM2\&MM3, we used the core catalogs in \citet{Motte2018-2,Pouteau2022-664,Nony2023-674,Louvet2024-690}. 
For the same observation in the ALMA-IMF survey \citep{Motte2022-662}, the core catalogs of W43-MM1\&MM2\&MM3 are presented in \citet{Pouteau2022-664,Nony2023-674,Louvet2024-690}. For W43-MM1, there is an independent catalog derived from the observation in \citet{Motte2018-2}. For the same region, different catalogs have different selection criteria and identifications, varying in the number of cores and their mass distribution, which is reflected in the differences in the CMF, as is shown in Fig.\ref{CMF}.

\section{Results and discussion}

\subsection{Observed CMFs}\label{o-cmf}

\begin{figure*}[htbp!]
\centering
\includegraphics[width=0.95\textwidth]{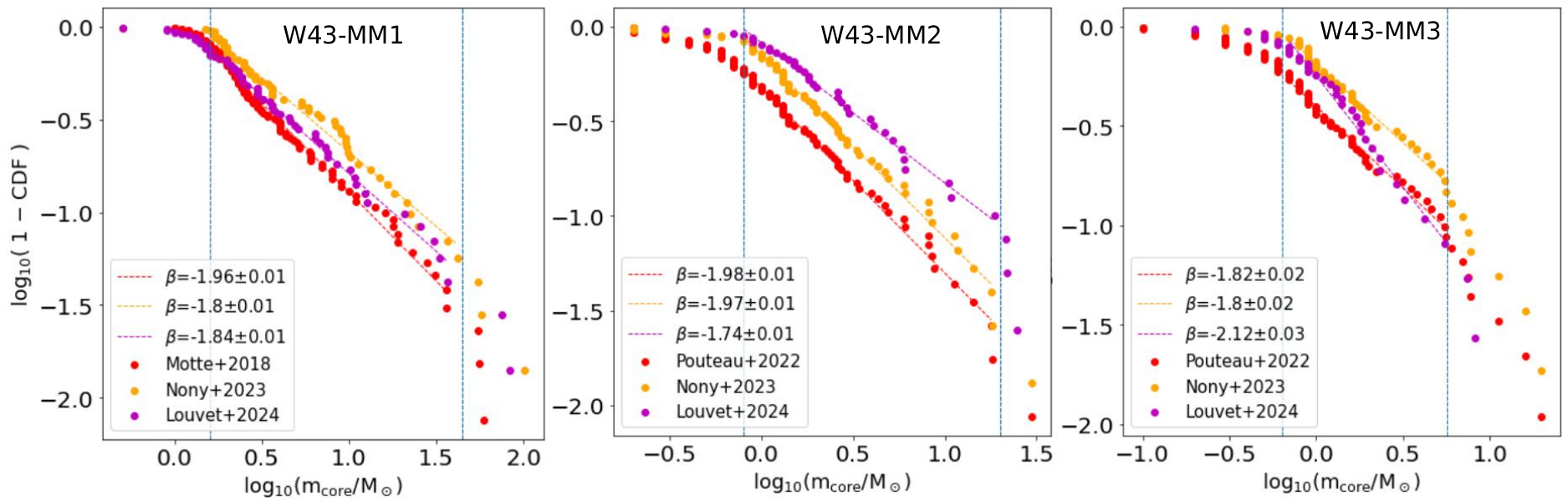}
\caption{Complementary cumulative distribution function (CCDF) plot and the linear fitting of the observed CMF in the mass range marked by two vertical dashed lines. $\beta$ is the slope of the corresponding $dN/dM$ mass function derived from linear fitting of the CCDF. The lower mass of the adopted fitting range corresponds to the observational mass completeness level estimated by \citet{Louvet2024-690}.}
\label{CMF}
\end{figure*}
 
For W43-MM2\&MM3, the core catalogs in \citet{Nony2023-674} were directly selected from those in \citet{Pouteau2022-664}, which only retains a portion of the original lists. The CMFs derived from the core catalogs in \citet{Pouteau2022-664} and \citet{Nony2023-674} have similar slopes, but the catalogs in \citet{Nony2023-674} with fewer cores shift systematically upward. For W43-MM1, three independent identifications in different observations give comparable CMFs. For W43-MM3, two independent identifications in the same observation give quite different CMFs.
Therefore, numerous observational constraints affect the CMF, including the calculation of the core's physical parameters, observational resolution, identification methods, and the completeness of the core population. However, the shallow nature of the CMF slope in W43 seems fairly robust at least.

\subsection{Clumps and embedded clusters}\label{mass}
\begin{table*}
\centering
\caption{Physical parameters of the corresponding ATLASGAL clumps.}
\begin{tabular}{ccccccccc}
\hline
region	&	clump	&	log$_{\rm 10}$($M_{\rm fwhm}$) ($M_{\odot}$)	&	$M_{\rm ecl}$ ($M_{\odot}$)	&	$m_{\rm max}$ ($M_{\odot}$)	& $N_{\rm star}$ \\
W43-MM3	&	AGAL030.718-00.082	&	3.173	&	399	&	26 & 798 \\
W43-MM2	&	AGAL030.703-00.067	&	3.211	&	435	&	27 & 870 \\
W43-MM1	&	AGAL030.818-00.056	&	3.248	&	473	&	29 & 946 \\
\hline
\label{clump}
\end{tabular}
\vspace{0mm}
\parbox{0.9 \linewidth}{ Notes: $M_{\rm fwhm}$ is the clump mass defined in \citet{Urquhart2022-510}. $M_{\rm ecl}$ is the mass of the embedded cluster finally formed in a clump. $m_{\rm max}$ is the mass of the most massive star in an embedded cluster. $N_{\rm star}$ is the total number of stars in a cluster.}
\end{table*}

Clumps are localized high-density structures within a molecular cloud and serve as the primary sites of star formation within the cloud \citep{Urquhart2018-473,Urquhart2022-510}. Cores are assumed to be overdense regions with respect to their parental clump.
Generally, a cloud includes many clumps, and a clump includes many cores. They appear as hierarchical, multiscale hub-filament structures \citep{Kumar2020-642,Zhou2022-514,Zhou2023-676,Zhou2024-686-146,Zhou2024PASA}.
To trace the evolution from the CMF to the IMF, the observation should cover the entire clump, as was done in the ALMA-IMF survey. A clump eventually forms an embedded cluster with an IMF \citep{Yan2017-607,Zhou2024PASP-1,Zhou2024PASP-2}. This IMF corresponds to the initial CMF in the clump. 
Therefore, clumps should be regarded as the fundamental units for studying the origin of the IMF \citep{Kroupa2024arXiv241007311K}.
W43 is a molecular cloud complex. The main star-forming regions in W43 are the clumps. The three subregions of W43 (W43-MM1\&MM2\&MM3) correspond to three independent clumps displayed in Table.\ref{clump}. They have independent star formation \citep{Zhou2024-688L}, and thus different CMFs, as is shown in Fig.\ref{CMF}.

The ATLASGAL clumps in \citet{Urquhart2022-510} corresponding to each region are listed in Table\ref{clump}. 
\citet{Zhou2024PASP-1,Zhou2024PASP-2} systematically studied the relationship between the physical parameters of the embedded clusters within ATLASGAL clumps with HII regions (HII-clumps) and the physical parameters of the clumps  themselves. We derived a $M_{\rm cl}-M_{\rm ecl}$ relation in \citet{Zhou2024PASP-2}:
\begin{multline}
\mathrm{log}_{10} (M_{\mathrm{cl}}/M_{\odot}) = 
(1.02 \pm 0.02) \times \mathrm{log}_{10} (M_{\mathrm{ecl}}/M_{\odot}) + (0.52 \pm 0.05).
\label{rl}
\end{multline}
Therefore, according to the clump mass ($M_{\rm cl}$), we can directly estimate the embedded cluster mass in stars ($M_{\rm ecl}$) in the clump. Although this relation is derived from HII-clumps, considering that the clumps in different evolutionary stages have similar mass distributions \citep{Urquhart2022-510}, it can also be used for younger clumps.

\subsection{Initial stellar populations}

Based on the recipe described in Sec.\ref{initial}, the initial stellar population in each high-mass star-forming region was generated using the updated \texttt{McLuster} program \citep{Kupper2011-417,Wang2019-484}, according to the embedded cluster mass ($M_{\rm ecl}$) estimated in Sec.\ref{mass}. Assuming that the average mass of stars in a cluster is 0.5 M$_{\odot}$, the total number of stars in a cluster ($N_{\rm star}$) is $M_{\rm ecl}$/(0.5 M$_{\odot}$). The proportion of binaries in the initial stellar population ($f_{\rm b}$) is an adjustable parameter in the \texttt{McLuster} code. Since the primary and secondary mass of the binary was selected randomly from the canonical IMF, for each embedded cluster, we ran \texttt{McLuster} 100 times, and overlapped these 100 stellar populations to derive a synthesized core population. For binaries, we summed the masses of each pair corresponding to one core. For single stars, we retained the generated mass. Then the stellar mass was divided by $\epsilon_{\rm core}$ to construct the mass distribution of the cores.

To derive the CMF, one common approach is to fit a linear relationship to data that has been grouped into evenly sized bins in logarithmic space. However, binning can lead to substantial information loss, as it aggregates data points into groups rather than using each one individually, which can obscure finer details in the distribution. This issue is especially pronounced in the higher mass range, where bins may contain very few data points, resulting in slope estimates that are often skewed and unreliable \citep{Maiz2005-629,Maschberger2009-395}. To address these limitations, \citet{Koen2006-365} employed a complementary cumulative distribution function (CCDF) plot, which avoids binning and retains more detail than traditional binning methods.
Taking $m_{\mathrm{max}} =$100 $M_{\odot}$, 
equation\ref{IMF} was used to generate mock data of the IMF.
According to the theoretical $m_{\rm max}$-$M_{\rm ecl}$ relation derived in \citet{Zhou2024PASP-2},
$m_{\mathrm{max}} = $100 $M_{\odot}$ corresponds to $M_{\rm ecl} \approx$ 5000 $M_{\odot}$; thus,
the number of stars ($N_{\rm star}$) in the realization is 10000.
Fig.\ref{IMF} shows the CCDF plot of the IMF. 

\subsection{Comparison of synthesized and observed CMFs}


\begin{figure*}[htbp!]
\centering
\includegraphics[width=1\textwidth]{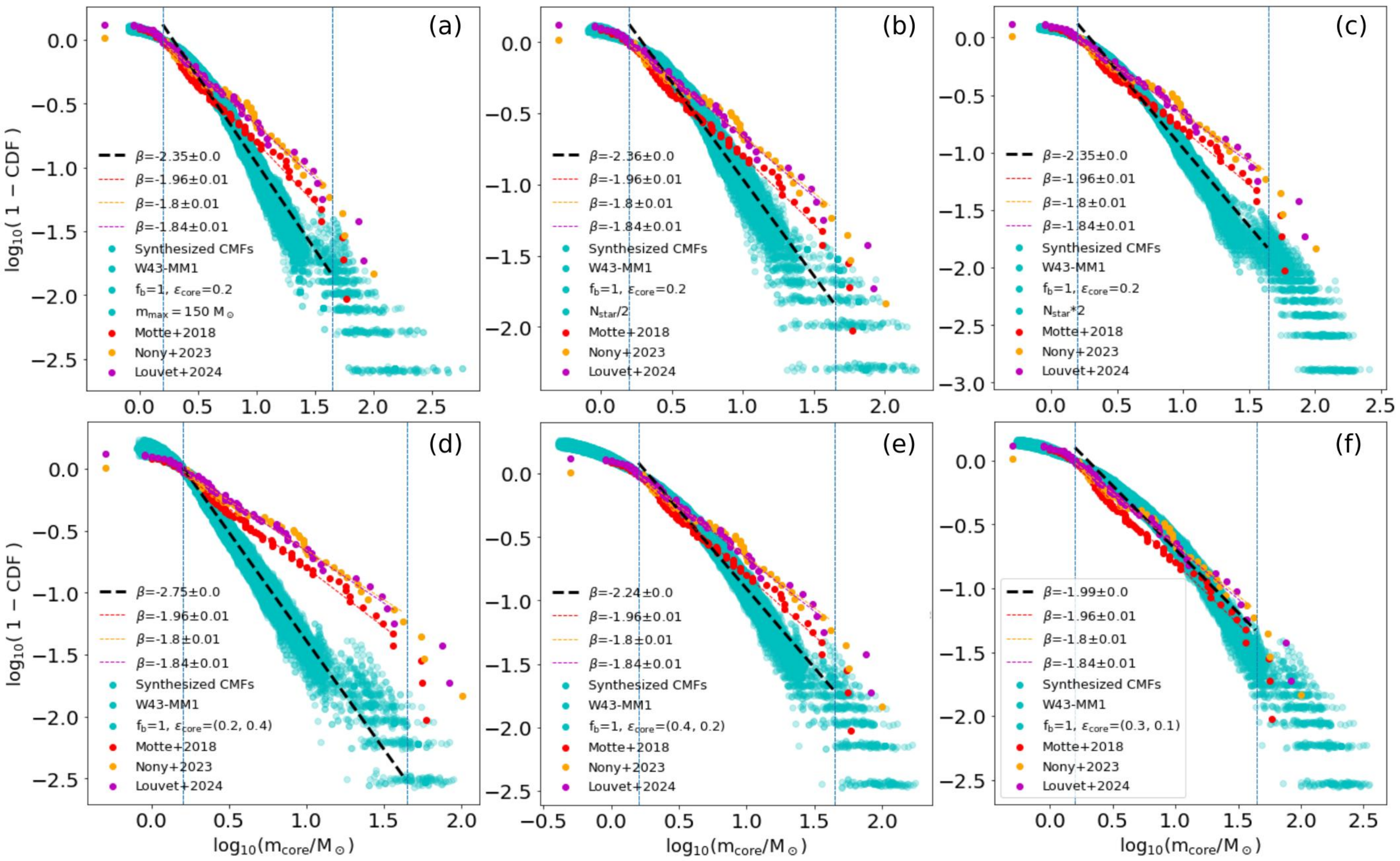}
\caption{Complementary cumulative distribution function (CCDF) plot and the linear fitting of the synthesized CMFs (in cyan) for W43-MM1 by setting $f_{\rm b}$=1 and $\epsilon_{\rm core}$=0.2. (a)-(c) The cases of $m_{\mathrm{max}}$=150 $M_{\odot}$, half the original number of stars ($N_{\rm star}$/2) and twice the original number of stars ($N_{\rm star}$*2), respectively; (d) The case in which $\epsilon_{\rm core}$ decreases with decreasing core mass. $\epsilon_{\rm core}$ was randomly selected from the range (0.2, 0.4); (e) The case in which $\epsilon_{\rm core}$ increases with decreasing core mass. $\epsilon_{\rm core}$ was randomly selected from the range (0.2, 0.4); (f) Same as panel (e), but $\epsilon_{\rm core}$ is in the range (0.1, 0.3). All selected $\epsilon_{\rm core}$ in panels (d), (e), and (f) are sorted in descending or ascending order. All CCDFs (both observed and synthesized) were normalized to 1 at the mass completeness level marked by the left vertical dashed line.}
\label{para1}
\end{figure*}

\begin{figure}
\centering
\includegraphics[width=0.4\textwidth]{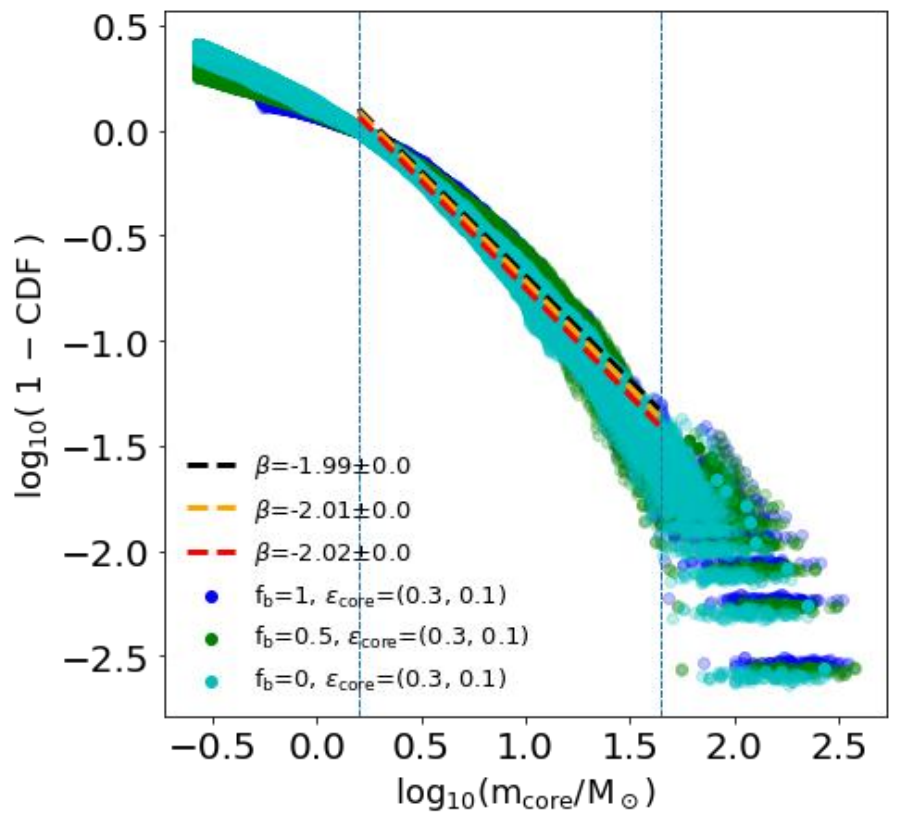}
\caption{Comparison of the synthesized CMFs for W43-MM1 for $f_{\rm b}$=1, 0.5 and 0. All synthesized CCDFs were normalized to 1 at the mass completeness level marked by the left vertical dashed line.}
\label{fb}
\end{figure}

\begin{figure}[htbp!]
\centering
\includegraphics[width=0.4\textwidth]{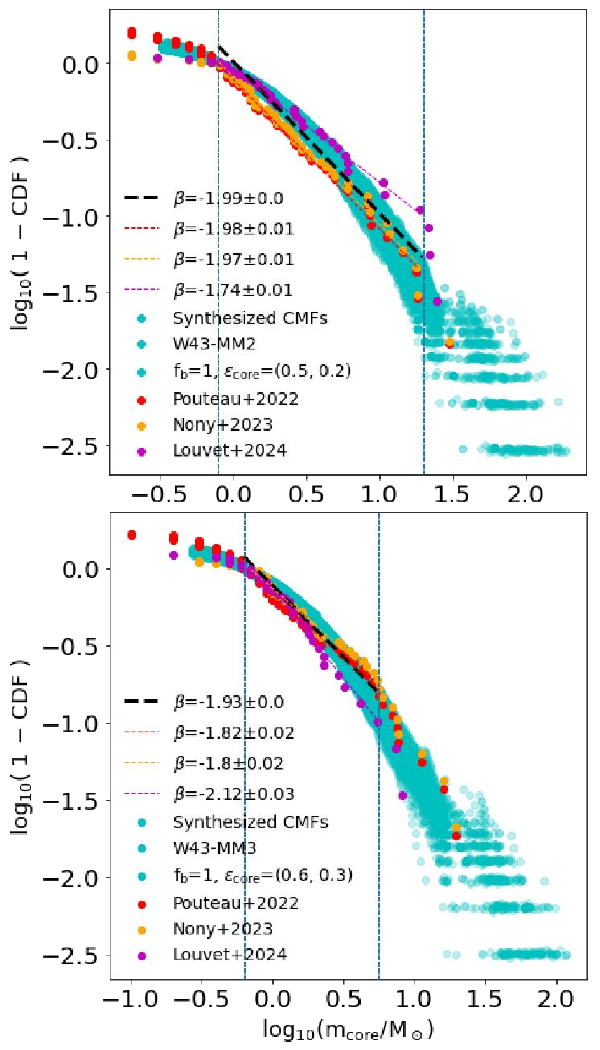}
\caption{Same as Fig.\ref{para1}(f) for W43-MM1. Here is the comparison of the synthesized CMFs and the observed CMFs for W43-MM2/MM3, assuming that $\epsilon_{\rm core}$ increases with decreasing core mass. All CCDFs (both observed and synthesized) were normalized to 1 at the mass completeness level marked by the left vertical dashed line.}
\label{para2}
\end{figure}

\begin{figure}
\centering
\includegraphics[width=0.35\textwidth]{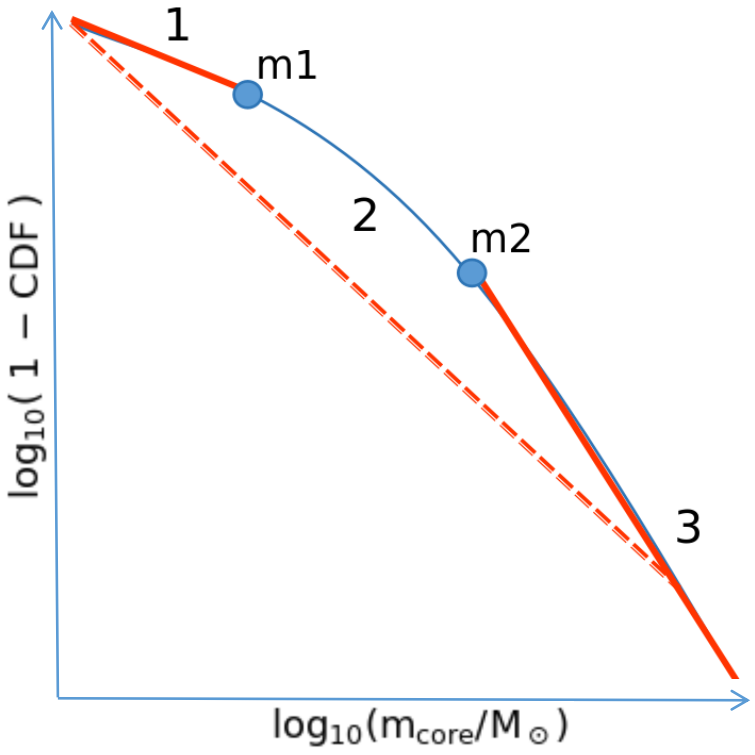}
\caption{The synthesized CMF can be roughly divided into three segments, i.e., “1,” “2,” and “3.” Segment “2” is the transition part. “m1” and “m2” mark the mass range of the transition. Segment “3” is the real CMF of high-mass cores. The dashed red line is the observed CMF, for which observational biases prevent one from distinguishing the transition, resulting in a combined fit of three different segments with a shallower slope. Thus, the observed top-heavy CMF is an artifact.}
\label{carton}
\end{figure}


For all three regions in Fig.\ref{CMF}, the low-mass end of the CMFs does not exhibit the clear transition seen in the IMF, which may be caused by the uncertainty in core mass estimation. As is discussed in \citet{Louvet2024-690}, the primary sources of error in the mass estimates stem from uncertainties in the opacity index and core temperature. To account for these uncertainties, they conducted a Monte Carlo simulation on 330 cores exceeding the 1.64 $M_{\odot}$ completeness threshold, allowing simultaneous variations in the opacity index, temperature, and source flux. Fig.6 of \citet{Louvet2024-690} presents the cumulative CMFs obtained from 10$^{3}$ simulation trials, where we can see a clear slope transition.

According to the observed CMF, we have fixed the mass range of the linear fitting for each clump in Fig.\ref{CMF}. For the synthesized core population of each clump, we strictly follow the fixed mass range to do linear fitting (the dashed black segments in Fig.\ref{para1}, Fig.\ref{fb} and Fig.\ref{para2}). 
A direct comparison between the synthesized and observed CMFs is only meaningful above the estimated mass completeness level of the observed sample. Therefore, for the purpose of direct comparison, all CCDFs (both observed and synthesized) were normalized to 1 at the mass completeness level marked by the left vertical dashed line in Fig.\ref{CMF}.

In Fig.\ref{para1}(f), 
there is a noticeable transition region in the synthesized CMFs, where the slope changes from flat to steep; that is, segment “2” in Fig.\ref{carton}. As is illustrated by the dashed red line in Fig.\ref{carton}, if the fitting range lies precisely within this transition region, it encompasses both the flat and steep slopes, resulting in an overall flat slope. If an observed CMF truly exhibits such a slope transition but observational biases obscure it, leading to a combined fit with a shallower slope, then the apparent top-heavy CMF would be merely an artifact.
To accurately fit the high-mass end of the CMF, it is essential to identify whether the critical masses for the transition -- “m1” and “m2” in Fig.\ref{carton} -- exist. If so, this critical mass, m2, should be used as the starting point for fitting the slope of the high-mass end, which can then be compared to the Salpeter value.
These issues appear in Fig.6 of \citet{Louvet2024-690},
in which a combined fit results in a shallower CMF slope. When focusing only on the high-mass end of this figure, the CMF slope is actually comparable to the Salpeter value.

By setting $f_{\rm b} =1$ and $\epsilon_{\rm core} =0.2$,
in Fig.\ref{para1}(a-c), we explored how the selected $m_{\rm max}$ and $N_{\rm star}$ (or $M_{\rm ecl}$) influence the synthesized CMFs, and found that the synthesized CMFs are not sensitive to these parameters. A larger $m_{\rm max}$ will increase the mass at the high-mass end, which is beyond the observed core mass range. For the same reason, we do not consider systems of a higher order than binaries (such as triples and quadruples) in this work.
In Fig.\ref{fb}, the synthesized CMFs with $f_{\rm b} =1$ do not provide better fits than those with $f_{\rm b} =0.5$ or $f_{\rm b} =0$. A high multiplicity fraction in the synthesis is not necessary. 


However, in Fig.\ref{para1}(a-c), the synthesized CMFs do not align with the observed ones. The reason may be that, in each case, we apply a constant $\epsilon_{\rm core}$ for all the synthesized cores. A likely possibility is that cores of different masses have different $\epsilon_{\rm core}$. In Fig.\ref{para1}(d-e), we compare two scenarios in which $\epsilon_{\rm core}$ either increases or decreases with decreasing core mass, respectively. In each case, $\epsilon_{\rm core}$ was randomly selected from a range, then, all selected $\epsilon_{\rm core}$ were sorted in descending or ascending order.
We found that in order to match the observed CMFs, $\epsilon_{\rm core}$ must increase with decreasing core mass. A possible explanation is that higher-mass cores contain more gas, leading to the formation of more massive stars, and stronger feedback ultimately disperses more gas. This is similar to the observational results at the clump scale \citep{Zhou2024PASP-2}, where the star formation efficiency (SFE) is strongly correlated with the clump mass: higher-mass clumps have a lower SFE. Due to observational resolution limitations, many of the so-called high-mass cores may actually be sub-clumps, which eventually form multiple systems, rather than just a single star. Furthermore, we should keep in mind that the cores in high-mass star formation regions can continue to accrete and accumulate mass during the process of star formation. These need to be considered when estimating $\epsilon_{\rm core}$.


To align the synthesized CMF with the observed CMF, we need to assume lower-mass cores that have higher $\epsilon_{\rm core}$, and then adjust the range of $\epsilon_{\rm core}$ accordingly. As is shown in Fig.\ref{para1}(f) and Fig.\ref{para2}, this approach successfully aligns the synthesized CMFs of W43-MM1/MM2/MM3 with the observed ones. 
Since the currently observed CMF may be not robust, we do not discuss the selection of optimal parameters further, as this depends on the specific CMF. Ideally, the mass dependence of $\epsilon_{\rm core}$ would be directly determined from observations in future work.

In this work, we have assumed a canonical IMF and inversely reconstructed the observed CMFs to investigate potential factors contributing to a top-heavy CMF. The factors discussed in this work may or may not be valid, implying that the top-heavy CMF could be genuine. The variation in the IMF with environmental conditions has been extensively studied \citep{Lee2009-706,Gunawardhana2011-415,Mor2019-624,Dib2023-959,Kroupa2024arXiv241007311K,del2025arXiv250117236D}. Rather than asserting the universality of the IMF, this paper emphasizes potential issues in the CMF fitting that warrant careful consideration.


\begin{acknowledgements}
We thank the referee for providing constructive review comments, which have helped to improve and clarify this work.
PK thanks the DAAD Eastern European Exchange Programme at Bonn and Charles University for support.
\end{acknowledgements}

\bibliographystyle{aa} 
\bibliography{ref}


\appendix

\section{Supplementary figures}

\begin{figure}[htbp!]
\centering
\includegraphics[width=0.4\textwidth]{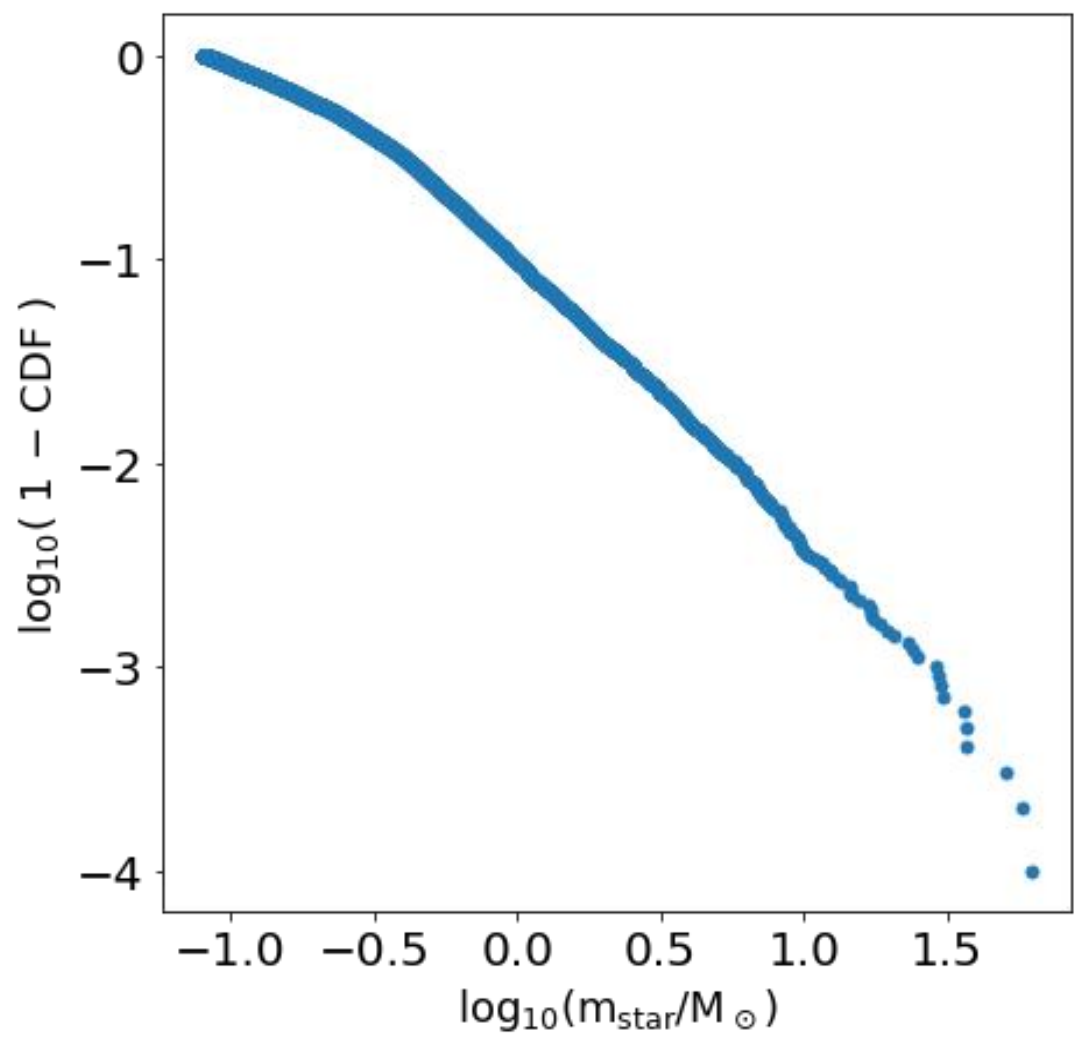}
\caption{Complementary cumulative distribution function (CCDF) plot of the IMF in equation.\ref{IMF} by taking $m_{\mathrm{max}} =$100 $M_{\odot}$ and $\alpha_3=2.3$.}
\label{IMF}
\end{figure}

\section{Initial stellar populations}\label{initial}

\citet{Kroupa1995b-277} proposed a model for the redistribution of energy and angular momentum within proto-binary systems, which directly results in binary characteristics such as mass ratio, eccentricity and period by the time the dynamical evolution of the star cluster begins to take effect. This process effectively generates the initial binary population (IBP), which serves as the starting point for star cluster simulations. The model has been validated against observational data and has successfully explained the binary properties observed in young clusters, associations, and Galactic field late-type binaries \citep{Kroupa2011-270,Marks2012-543}. 
The process of transforming birth binaries into initial binaries is commonly referred to as pre-main-sequence eigenevolution \citep{Kroupa1995b-277}, as it takes place during the pre-main-sequence phase of binary stars. The term “birth” refers to all protostars still embedded in circumprotostellar material. In contrast, “initial” refers to pre-main-sequence stars that are no longer embedded in such material, as it has been either accreted or expelled during the redistribution of energy and angular momentum. Over a period of approximately $10^5$ years, this pre-main-sequence eigenevolution converts the birth population into the initial population. \citet{Belloni2017-471} presents an updated version of pre-main-sequence eigenevolution, with the primary difference being the mass ratio distribution in the initial binary populations.

To undergo pre-main-sequence eigenevolution, binaries at birth are formed with specific distributions. The primary mass is selected randomly from the canonical IMF \citep{Kroupa2001-322}:
\begin{equation}
\label{IMF}
\xi_{\mathrm{\star}}(m) =
\begin{cases} 
0, & m<0.08~\mathrm{M}_{\odot}, \\
2k_{\mathrm{\star}} m^{-1.3}, & 0.08~\mathrm{M}_{\odot} \leqslant m<0.5~\mathrm{M}_{\odot}, \\ 
k_{\mathrm{\star}} m^{-2.3}, & 0.5~\mathrm{M}_{\odot} \leqslant m<1\mathrm{M}_{\odot}, \\
k_{\mathrm{\star}} m^{-\alpha_3}, & 1~\mathrm{M}_{\odot} \leqslant m<m_{\mathrm{max}}, \\
0, & m_{\mathrm{max}} \leqslant m,
\end{cases}
\end{equation}
where $\alpha_3=2.3$ is the constant Salpeter-Massey index for the invariant canonical IMF but will change for larger $\rho_{\mathrm{cl}}$ (the density of star-forming clump) to account for IMF variation under star-burst conditions \citep{Elmegreen2003-338, Dib2007-381,Dabringhausen2009-394,Marks2012-422,Dib2023-959,Kroupa2024arXiv241007311K}. As verified in \citet{Zhou2024PASP-2}, we take $\alpha_3=2.3$ in this work. 
The value of $m_{\mathrm{max}}$ is decided by the theoretical $m_{\rm max}-M_{\rm ecl}$ relation \citep{Yan2023-670,Zhou2024PASP-2}. The secondary mass is also randomly drawn from the same IMF, meaning the binary components are paired randomly. It’s important to note that the primary and secondary stars are distinguished only after both stars have been independently chosen from the canonical IMF.

In \citet{Kroupa1995b-277}, the pre-main-sequence eigenevolution theory was developed mainly to explain observed properties of Galactic field low-mass binaries (A, F, G, K and M-dwarfs). We note that low-mass binaries in this work correspond to all binaries whose primary mass is smaller than $5$ M$_\odot$. Binaries with primary masses greater than this are called high-mass binaries. For high-mass binaries with O and B-dwarfs as primaries (greater than $5$ M$_\odot$), we utilize observational distributions derived from O and B-dwarfs in \citet{Sana2012-337} to generate the birth population. Here we assume that the process of transforming the birth population into the initial population for high-mass binaries has already occurred, and directly utilize observational distributions, since this transformation occurs rapidly for massive stars. The mass ratio distribution of high-mass binaries is a uniform distribution \citep{Oh2015-805,Belloni2017-471}. The eccentricity and period distributions of both low- and high-mass binaries are described in \citet{Belloni2017-471}.

\end{document}